\documentclass[twocolumn,showpacs,prl]{revtex4}
\usepackage{amssymb}

\usepackage{amsmath}
\usepackage{epsf}
\usepackage{graphicx}
\usepackage{dcolumn}
\usepackage{bm}

\begin{document}

\title{Delocalization by Disorder in Layered Systems}
\author{Dmitrii L. Maslov$^{a}$, Vladimir I. Yudson$^{b}$, Andres M. Somoza$^{c}$,
and Miguel Ortu\~{n}o$^{c}$}
\date{\today}

\affiliation{
$^{a}$Department of Physics, University of
Florida, P. O. Box 118440, Gainesville, FL
32611-8440\\
 $^b$Institute for
Spectroscopy, Russian Academy of Sciences, Troitsk, Moscow region,
142190, Russia\\
$^c$Departamento de F\'{\i}sica-CIOyN, Universidad de
Murcia, Murcia 30.071, Spain}

\begin{abstract}

Motivated by anomalously large conductivity anisotropy in layered materials,
%
%
we propose a simple model of randomly spaced potential barriers (mimicking stacking
faults) with isotropic impurities in between the barriers. We solve this model both
numerically and analytically, by utilizing an exact solution for the conductivity of
a one-dimensional (1D) disordered system. In the absence of bulk disorder, electron motion in the out-of-plane direction is localized. Bulk disorder destroys 1D localization. As a result, the out-of-plane conductivity
is finite and scales linearly with the scattering rate by bulk impurities until
planar and bulk disorder become comparable. The \emph{ac} out-of-plane conductivity
is of a manifestly non-Drude form, with a maximum at the frequency corresponding to the
scattering rate by potential barriers.

\end{abstract}
\maketitle

It is usually the case that cleaner metals are better conductors. In the
semiclassical, phase-incoherent regime of transport, this happens simply because
stronger disorder means a shorter scattering time; in the phase-coherent regime,
stronger disorder enhances Anderson localization thereby reducing the conductivity even further. 
It is also commonly believed that localization can be
destroyed only by inelastic scattering. In this Letter, we propose and analyze a
simple model with \emph{two} types of disorder which defies these notions. We show
that an increase in \emph{one }type of disorder leads to a destruction of the
Anderson-localized state and, consequently, to a \emph{increase} in the conductivity
in one direction.

\begin{figure}[tbh]
\includegraphics[width=0.6\textwidth]{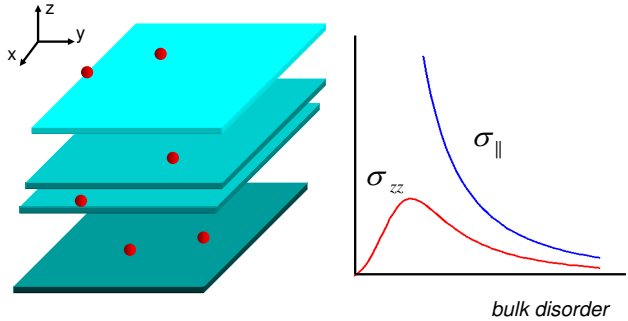}
\caption{(Color online) Left: a system of randomly spaced parallel potential barriers
and randomly distributed isotropic impurities. Right: expected dependences of the
in- and out-of-plane conductivities on bulk disorder.} \label{fig:cartoon}
\end{figure}

The model consists of planar barriers located at random spacings to each other
\emph{and} isotropic impurities distributed randomly in between the barriers (see
Fig. \ref{fig:cartoon}).
This model is motivated by some well-known but hitherto unexplained peculiarities of
electron transport in layered conductors. In the band picture, the conductivity in a
certain crystallographic direction scales with the inverse effective mass in
this direction. In many cases, however, the observed ratio of the in-plane and
out-of-plane conductivities exceeds the (inverse) ratio of the effective masses by
several orders of magnitude. A well-known case of such an anomaly is graphite,
where the conductivity ratio exceeds the mass ratio by 2-3 orders of magnitude \cite
{graphite}, but other materials, e.g., NaCo$_{2}$O$_{4}$ \cite{singh}, cuprates
\cite{pickett}, etc., also provide examples of this behavior. Stacking faults, e.g.,
''wrong'' planes violating Bernal stacking of graphene sheets in graphite, have
been proposed to be responsible for abnormally large conductivity anisotropy long time
ago \cite{ono};
however, little attention has been paid to localization of electrons by an
\emph{array} of faults.

We consider a system of electrons with separable but otherwise arbitrary
spectrum 
$\varepsilon (\vec{k}_{||},k_{z})=\varepsilon _{||}(\vec{k}%
_{||})+\varepsilon _{z}(k_{z})$, 
subject to two types of random potential: the 1D potential of the barriers, $%
U\left( z\right)$, and the 3D potential of isotropic impurities, $V\left(
\vec{r}\right)$.
In the absence of bulk disorder, the in- and out-of-planes degrees of
freedom separate. Accordingly, the electron wave function is factorized as
$\Psi \left( \vec{r}_{||},z\right) =\varphi
\left( \vec{r}_{||}\right) \chi \left( z\right)$, with $\chi(z)$ satisfying
an effectively 1D Schroedinger equation
$\left[ \varepsilon _{z}(-i\partial _{z})+U\left( z\right) \right] \chi
\left( z\right) =\left( E-\varepsilon _{||}\left( \vec{k}_{||}\right)
\right) \chi \left( z\right)$,
where $\vec{k}_{||}$ is the (quasi) momentum along the planes.
All states of such a system are localized in the $z$-direction by infinitesimally
weak disorder. Therefore, the \emph{dc }conductivity across the planes, $%
\sigma _{zz},$ is zero. On the other hand, since barriers do not affect the electron motion
along the planes, the in-plane conductivity, $\sigma _{||} $, is infinite. Bulk
disorder mixes the in- and out-of-planes degrees of freedom, so that the separation
of variables is no longer possible. Therefore, 1D localization in the $z$ direction
is destroyed, and $\sigma _{zz}$ increases with bulk disorder, as long as it remains
weaker than the planar one. When two disorders become comparable, $\sigma _{zz}$
reaches a maximum and decreases upon a further increase in bulk disorder in accord
with the Drude formula \cite{comment}.  At the same time, $\sigma _{||}$
decreases monotonously with bulk disorder. A sketch of expected dependences of
$\sigma _{zz}$ and $\sigma _{||}$ on 3D disorder is presented in
Fig.~\ref{fig:cartoon} (right). In the rest of the paper, we confirm this simple picture both
numerically, by calculating $\sigma _{zz}$ in the Anderson model, and analytically,
by exploiting the Berezinskii solution of the 1D localization problem.


Numerically, we study the Anderson model with nearest-neighbor hopping (set
to unity to fix the energy scale) on a cubic lattice (of unit spacing)
\begin{equation}
H=-\sum_{{\mathbf i},\mathbf{j}}a_{\mathbf{j}}^{\dagger }a
_{\mathbf{i}}+\sum_{\mathbf{i}}\epsilon _{\mathbf{i}}
a_{\mathbf{i}}^{\dagger
}a_{\mathbf{i}}
+\mathrm{H.c.}\;
\end{equation}
Here, the on-site energy $\epsilon _{\mathbf{i}}=\phi _{\mathbf{i}}+\eta _{i_z}$ and $\mathbf{i} = (i_x,i_y,i_z)$.
The first term, $\phi _{\mathbf{i}}$, is the standard (bulk) disorder term which is chosen
independently for each site in the interval $(-W_{\mathrm{B}}/2,W_{\mathrm {B}}/2)$ with uniform probability.
The second term, $\eta _{i_z},$ describing planar disorder, is chosen as $-W$ with
probability $p$ and as $W$ with probability $1-p$. For all results reported in this
paper, $p=1/2$. The simulations are done at the energy equal to 0.1, to avoid the
center of the band. We employ the recursive Green's function technique \cite
{M85} with periodic boundary conditions in the directions transverse to the $%
z$-axis. The out-of-plane conductance $G_{zz}$ is equal to ${2e^{2}}T/h$, where $T$
is the transmission coefficient between two wide leads. 
 The simulations were performed for cubic samples of sizes $L$ up to 35
lattice spacings. The bandwidths of planar disorder $W$ were chosen as $1,1.5,2,2.5$
and $3$, which corresponds to localization lengths between roughly 2 and 15 lattice
spacings, in the absence of bulk disorder. The bandwidth of bulk disorder $W_{\mathrm{B}}$ ranged
in between $0$ and $18$. We have averaged $\log G_{zz}$ for $10^{3}$ samples for each
set of parameters. Crystalline anisotropy can be simply accounted for in simulations; however,
the conductance is already anisotropic due to anisotropy of disorder even on a cubic lattice.

\begin{figure}[tbh]
\includegraphics[width=.48\textwidth]{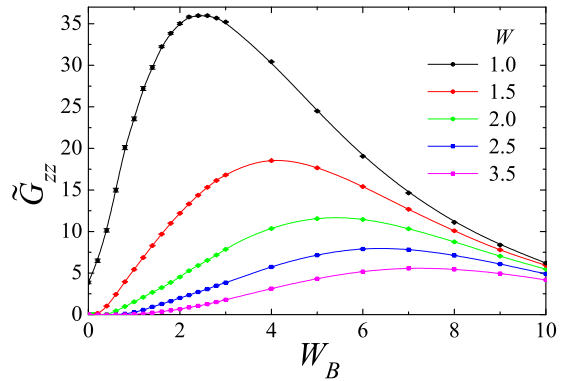}
\caption{(Color online) Out-of-plane conductance versus the bandwidth of
bulk disorder $W_{\mathrm{B}}$ for a range of values
of planar disorder $W$, as shown in the figure, and $L=30$.}
\label{fig:GvsB}
\end{figure}

Figure \ref{fig:GvsB} shows $\tilde{G}_{zz}\equiv \exp \{\langle \ln G_{zz}\rangle
\}$ as a function of bulk disorder for several values of planar disorder. As
expected,
an increase in bulk disorder leads first to an in increase in $\tilde{G}_{zz}$ followed by a subsequent decrease. The
position of the peak depends on planar disorder but is almost independent
of $L$. We checked that the conductance scales
linearly with $L$ for most of the range
of parameters represented in Fig.\ 2, so that we are in the diffusive regime. Specifically, the diffusive regime begin when
the conductance becomes larger than $2e^{2}/h$ and continue up to the 3D Anderson
transition (not shown in Fig.~\ref{fig:GvsB}).

\begin{figure}[tbh]
\includegraphics[width=.48\textwidth]{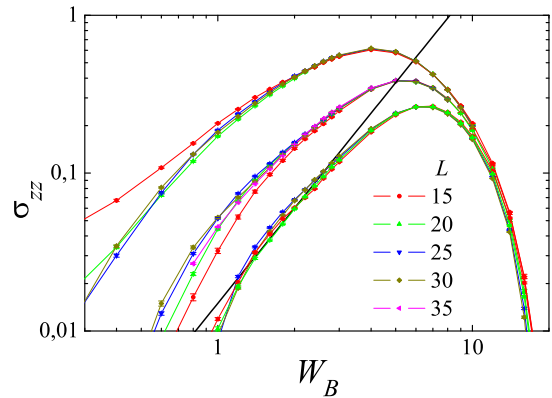}
\caption{(Color online) Out-of-plane conductivity versus the bandwidth of bulk
disorder $W_{\mathrm{B}}$ on a double logarithmic scale for a range of system sizes, as shown in
the figure, and three values of planar disorder: $W=1.5$ (upper set), $W=2$ (middle
set), and $W=2.5$ (lower set).} \label{fig:logGvsB}
\end{figure}

Figure \ref{fig:logGvsB} shows the collapse of the data for the
conductivity, $\sigma _{zz}=\tilde{G}_{zz}/L$, on a double-logarithmic plot.
Three sets of curves corresponds to three values of planar disorder: $%
W=1.5 $ (upper set), $W=2$ (middle set) and $W=$2.5 (lower set). Within each
set, the conductivity was computed for 
different values of $L,$ as
indicated in the legend.
The
straight line has a slope equal to two. This scaling is confirmed by the
analytic solution of the model, described below.

To solve the problem analytically, we adopt the delta-correlated forms for both types
of disorder $\langle U\left( z\right) U(0)\rangle =\gamma _{z}\delta \left( z\right)
$ and $\langle V\left( \vec{r}\right) V\left(0\right) \rangle =\gamma \delta
\left( \vec{r}\right) $, and assume that bulk disorder is weaker than planar one,
i.e., $1/\tau \equiv $ $2\pi \nu _{3}(E_{F})\gamma \ll 1/\tau _{z}=2\pi \nu
_{1}(E,\vec{k}_{||})\gamma _{z},$ while planar disorder is weak in a sense that
$E_{F}\tau _{z}\gg 1.$ Here, $\nu _{3}$ is the 3D density of states and $\nu _{1}$ is
the 1D density of states at fixed value of $\vec{k}%
_{||}$ per one spin orientation. In the absence of bulk disorder, our problem reduces to
the 1D case with the velocity $v_{z}=|\partial \varepsilon (\vec{k}%
_{||},k_{z})/\partial k_{z}|_{k_{z}=k_{zF}(\vec{k}_{||})} $, where
$k_{zF}(\vec{k}_{||})$ is a positive root of the equation $\varepsilon
(\vec{k}_{||},k_{z})=E_{F},$ and the scattering time $\tau _{z}$ being functions of
$\vec{k}_{||}$. The result for the $ac$ conductivity of a strictly 1D disordered
system, surmised first by Mott \cite{Mott} and derived rigorously by Berezinskii
\cite{Berezinskii} reads
\begin{equation}
\hspace{-0.cm} \sigma ^{1D}(\omega )=\frac{16e^{2}v_{z}\tau _{z}}{\pi }\left[ -i\zeta
(3)\omega \tau _{z}+\, 2\tau _{z}^{2}\omega ^{2}\ln ^{2}{(\omega \tau _{z})}%
\right], \label{berez} \hspace{-0.2cm}
\end{equation}
for $\omega \tau _{z}\ll 1$. (The numerical coefficient in the imaginary part was
corrected in Refs.~\cite{Gogolin-Melnikov-Rashba,Abrikosov&Ryzhkin}). The out-of-plane
conductivity of a 3D sample with $V=0$ is obtained from Eq. ~(\ref{berez}) by summing 
over $\vec{%
k}_{||}$: $\sigma _{zz}\left( \omega \right) =\int d^{2}k_{||}\sigma ^{1D}\left(
\omega \right)/(2\pi )^{2} $. As expected, $\sigma _{zz}\left( 0\right) =0.$

\begin{figure}[tbh]
\includegraphics[width=0.48\textwidth]{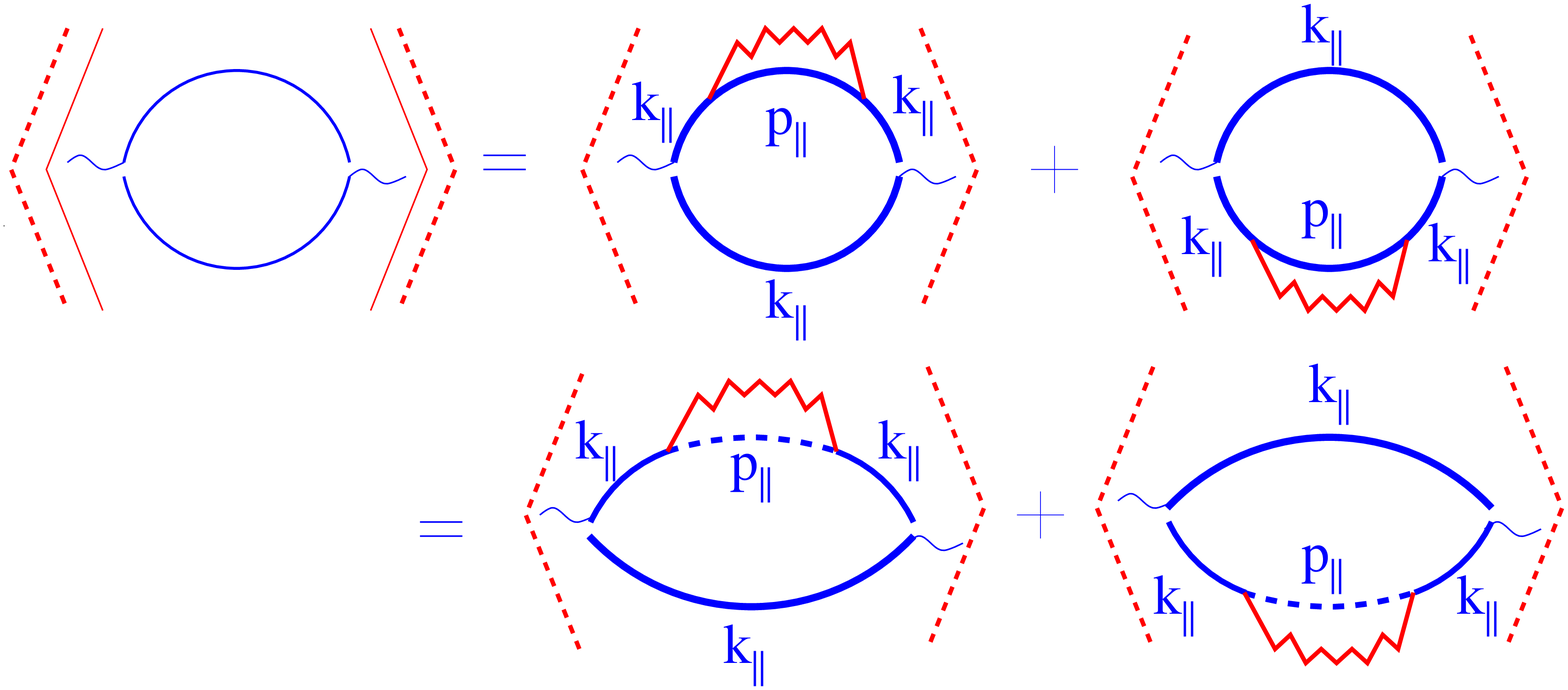}
\caption{(Color online) Diagrams for the out-of-plane conductivity to leading order in
bulk disorder. Thin lines: exact Green's functions in the presence of both types of
disorder; thick solid lines: Green's functions in the presence of planar disorder
only; thick dashed lines: Green's functions averaged over planar disorder; zigzag:
correlator of bulk disorder; solid and dashed brackets: averaging over planar and
bulk disorder, correspondingly.} \label{fig:diagram}
\end{figure}

In the presence of both types of disorder, $\sigma _{zz}$ is given by the Kubo formula
\begin{eqnarray}%
\label{sigma-zz}
&&\sigma_{zz}(\omega)= \frac{e^2}{2\pi} \frac{1}{\mathcal{A}%
^3}\sum_{\vec{k}_{||}, \, \vec{k}'_{||}} \int dz' \nonumber \\
&& \times\langle \langle v_z
\mathcal{G}^R_{+}(\vec{k}_{||},z; \vec{k}'_{||}, z') v'_z\mathcal{%
G}^A_{-}(\vec{k}'_{||},z'; \vec{k}_{||}, z)\rangle_{\mathrm{p}} \rangle_{\mathrm{b}}
\, ,
\end{eqnarray}
where $\mathcal{G}_{\pm }^{R(A)}=\mathcal{G}^{R(A)}(\vec{k}_{||},z;\vec{k}%
_{||}^{\prime },z^{\prime };E_F\pm \omega /2)$ is an exact retarded (advanced)
electron Green's function in the mixed $\vec{k}_{||}-z$ representation for a given
disorder realization, $\mathcal{A}$ is the sample area in the lateral direction, and
$\langle \dots \rangle _{\mathrm{b,p}}$ denotes averaging over bulk and planar
disorders, correspondingly. The diagram for $\sigma _{zz}$ is shown in Fig.
\ref{fig:diagram} on the left. To leading order in $\gamma$, the conductivity
$\sigma^{(1)} _{zz}$ averaged over bulk disorder is given by the sum of the two
diagrams in the first row of Fig.~\ref{fig:diagram}, where thick solid lines denote
Green's functions in the
absence of bulk disorder, $\mathcal{G}_{\mathrm{p}}^{R(A)}(z,z^{\prime };\vec{%
k}_{||};E),$ and zigzags denote the correlation function of bulk disorder. There are
no vertex corrections for the case of delta-correlated bulk disorder. The first
(second) diagram in the
first row
of Fig.~\ref{fig:diagram} is obtained by
replacing the 
exact Green's function
 by
$\gamma\int_{z_1,\vec{p}_{||}}
\mathcal{G}_{\mathrm{p}}^{R(A)}(z,z_1;\vec{%
k}_{||})
\mathcal{G}_{\mathrm{p}}^{R(A)}(z_1,z_1;\vec{p}_{||})
\mathcal{G}_{\mathrm{p}}^{R(A)}(z_1,z^{\prime};\vec{%
k}_{||})$.

Subsequent averaging over planar disorder is simplified dramatically by
noticing that the effective energies $E-\varepsilon _{||}(\vec{k}_{||})$ of
the Green's functions depend on a particular
value of $\vec{k}_{||}$. For short-range bulk disorder, the momentum $%
\vec{p}_{||}$ of the Green's function below the zigzag line differs considerably from
the momentum $\vec{k}_{||}$ in the rest of the diagram. This means that the typical
difference of corresponding energies is of order $E_{F}$, i.e., much greater than
$1/\tau _{z}$. In this situation, one can safely neglect correlations between the
Green's functions with different momenta and average
$\mathcal{G}_{\mathrm{p}}(z_{1},z_{1};\vec{p}_{||},E)$ over planar disorder
independently from the rest of the diagram. As a result, we arrive at the diagrams in
the second row of Fig.~\ref{fig:diagram}, where thick dashed lines denote the Green's
function averaged over planar disorder.
For weak planar
disorder ($E_{F}\tau _{z}\gg 1$), this Green's function is $\langle \mathcal{%
G}_{\mathrm{p}}^{R,A}\left( z,z;\vec{k}_{||};E\right) \rangle _{\mathrm{p}%
}=\int dk_{z}\left[ E-\varepsilon _{z}\left( k_{z}\right) -\varepsilon _{||}\left(
\vec{k}_{||}\right) \pm i/2\tau _{z}\right] ^{-1}/2\pi$ and the corresponding
self-energy insertion reduces to
$\Sigma ^{R(A)}(z,z^{\prime };\vec{k}_{||};E)=\mp \left(i/2\tau\right)\delta(z-z')$.
%
Expanding $\mathcal{G}^{R,A}_{\mathrm{p}%
}$ over the basis exact of eigenstates of the 1D problem, we reduce the convolution of two
Green's functions, sharing the point $z_1$, to
%
%
$\int_{z_{1}}\mathcal{G}_{\mathrm{p}}^{R,A}\left( z,z_{1};\vec{k}%
_{||};E\right) \mathcal{G}_{\mathrm{p}}^{R,A}\left( z_{1},z^{\prime
};\vec{k}_{||};E\right) =-\frac{\partial }{\partial E }\mathcal{%
G}_{\mathrm{p}}^{R,A}\left( z,z^{\prime };\vec{k}_{||};E \right)$.
Consequently, $\sigma _{zz}^{(1)}\left( \omega \right) $ is obtained from the exact
1D result via
\begin{equation}
\sigma _{zz}^{(1)}\left( \omega \right) =\frac{i}{\tau }\int \frac{%
d^{2}k_{||}}{\left( 2\pi \right) ^{2}}\frac{\partial \sigma ^{1D}\left(
\omega \right) }{\partial \omega }.  \label{first_gamma}
\end{equation}
To obtain\ the \emph{dc }conductivity,
one needs to differentiate only the imaginary part of Eq.~(\ref{berez}). This gives
\begin{equation}
\sigma _{zz}^{(1)}(0)=2e^{2}\nu _{3}(E_{F})D_{zz}\,,  \label{a2}
\end{equation}
where
\begin{equation}
D_{zz}=16\zeta (3)\frac{\langle v_{z}^{4}(\vec{k}_{||})\rangle _{||}%
}{v_{z,\max }^{4}}\frac{l_{z,\max }^{2}}{\tau }\,.  \label{a4}
\end{equation}
Here, $l_{z,\max }$ and $v_{z,\max }$ denote the maximum
values of $l_{z}(\vec{k}_{||})\equiv v_{z}\tau _{z}$ and $v_{z}(\vec{k}_{||})
$ attained for $\vec{k}_{||}=0$ and
$\langle f(\vec{k}_{||})\rangle _{||}=
\left(4\pi^2\nu _{3}(E_{F})\right)^{-1}
\int
d^{2}k_{||}
\nu _{1}(E_F,\vec{k}_{||})f(\vec{k}_{||})\,.$
The diffusion coefficient $D_{zz}$ is
proportional to the ratio of the square of the localization length in the 1D system
to the bulk scattering time. Numerically, we have found that $\sigma _{zz}(0)$ scales
as the square of the bulk disorder bandwidth. This
is confirmed by our analytic result since $\sigma
_{zz}^{(1)}(0)\propto 1/\tau \propto \gamma \propto B^{2}$ in the Born approximation.

Equation (\ref{a4}) allows for a simple physical interpretation. Bulk scattering
weakly couples 1D channels of localized electrons with different $\vec{k}_{||}$. Each
scattering event results in a random displacement of order $l_{z}$ in the $%
z$-direction and results in diffusion with the coefficient $D_{zz}\sim l_{z}^{2}/\tau
$. Notice that bulk disorder acts very similarly to the electron-phonon (e-ph)
interaction in a strictly 1D system, where $\sigma^{1D}(0)\propto
1/\tau_{\mathrm{e-ph}}$ \cite{Gogolin-Melnikov-Rashba}. The difference between the
two cases is that $\sigma^{1D}(0)$ scales with $1/\tau_{\mathrm{e-ph}}$ only at
temperatures higher than the single-level spacing within the localization length,
i.e, for $T\tau_z\gg 1$, while at lower temperatures $\sigma^{1D}$ is of the hopping
form. The condition $T\tau_z\gg 1$ allows one to neglect correlations between the
Green's functions in the self-energy insertions and in the rest of the diagram. In
our case, these correlations can be always neglected for short-range bulk disorder,
i.e., in contrast to phonon-activated transport, there is no \lq\lq hopping\rq\rq\,
regime for disorder-activated transport.

Coming back to the issue of anomalously large conductivity anisotropy, it is easy to
show that the in-plane conductivity is given by the usual Drude formula
$\sigma_{\alpha\beta} = \delta_{\alpha\beta}e^2\nu_3(E_F)\langle
v_{\alpha}v_{\beta}\rangle_{||}\tau$. Then the conductivity ratio can be estimated as
\begin{equation}
\sigma_{||}/\sigma_{zz}\sim
\left(\langle v^2_{||}\rangle_{||}/\langle v_{z}^2\rangle_{||}\right)
\left(\tau/\tau_{z,\max}\right)^2.
\end{equation}
As an example, we consider the case of graphite with $\sigma_{||}/\sigma_{zz}=
10^{4}$ at low temperatures. A realistic band structure model of graphite
\cite{graphite} gives $\langle v_{||}^2\rangle/\langle v_{z}^2\rangle\sim 140$, thus
$\tau_z/\tau\sim 0.12$. Taking $\tau=0.5\times 10^{-12}$\,s from Ref.~\cite{xudu} and
estimating $\langle v_z^2\rangle^{1/2}\sim 2\times 10^6$ cm/s, we obtain for the mean
free path due to planar disorder (stalking faults) in $l_z\sim 120$ \AA. This means
that stalking faults are separated by about a hundred perfect planes, which is quite
a realistic assumption.

Summing up higher-order diagrams with self-energy insertions due to bulk disorder
amounts to replacing the exact Green's functions in Eq. (\ref{sigma-zz})
by $\mathcal{G}^{R(A)}(z,z^{\prime };\vec{k}_{||};E\pm\omega/2\pm \frac{i}{2\tau }%
)$, which can be viewed as functions of a complex frequency. One can verify that all
intermediate steps in Refs.~\cite{Berezinskii} and \cite{Abrikosov&Ryzhkin} are valid
for complex $\omega$ as well. Therefore, the general result for the conductivity of
our model is obtained from the Berezinskii's solution as
\begin{equation}
\sigma _{zz}\left( \omega \right) =\int \frac{d^{2}k_{||}}{\left( 2\pi
\right) ^{2}}\sigma ^{1D}\left( \omega +\frac{i}{\tau }\right) .
\label{ac}\end{equation}
To lowest order in $%
1/\tau$, Eq.~(\ref{ac}) reduces back to Eq.~(\ref{first_gamma}).
Within the logarithmic
accuracy of the original Berezinskii's formula, we obtain
\begin{widetext}
\begin{subequations}
\begin{eqnarray}
\text{Re}\sigma _{zz}\left( \omega \right)  &=&2e^{2}\nu _{3}\left\langle
\frac{16l_{z}^{2}}{\tau }\left[ \zeta \left( 3\right) +\frac{2\tau _{z}}{\tau }%
\left( \omega ^{2}\tau ^{2}-1\right) \ln ^{2}\left\{ \omega ^{2}\tau
_{z}^{2}+\frac{\tau _{z}^{2}}{\tau ^{2}}\right\} \right] \right\rangle _{||}
\label{resigma}\\
\text{Im}\sigma _{zz}\left( \omega \right)  &=&-2e^{2}\nu _{3}\omega\tau\left\langle
\frac{16l_{z}^{2}}{\tau }\left[\zeta \left( 3\right)-\frac{2\tau_z}{\tau}\ln ^{2}
\left\{ \omega ^{2}\tau _{z}^{2}+\frac{\tau _{z}^{2}}{\tau ^{2}}\right\} \right]
\right\rangle _{||}.\label{imsigma}
\end{eqnarray}
\end{subequations}
\end{widetext}
These formulas are valid for an arbitrary value $\omega\tau $ but only for $\omega
\tau _{z}\ll 1$ and $\tau _{z}/\tau \ll 1. $ From Eq.~(\ref{resigma}), we see that
 Re$\sigma_{zz}(\omega)$ is almost constant for $\omega\ll\omega_{\mathrm{cr}}
\equiv 1/\left(\tau_{z,\max}\tau\right)^{1/2}$ and increases with $\omega$ in a Mott
way, as $\omega^2\ln^2\omega$, for $\omega\gg \omega_{\mathrm{cr}}$. At higher
frequencies, $\omega\gg 1/\tau_z$, $\sigma_{zz}(\omega)$  can be found perturbatively
in $1/\tau_z$: the leading order result is simply a Drude formula
$\sigma_{zz}(\omega)\propto 1/\omega^2\tau_z-i/\omega$. Therefore,  both
Re$\sigma_{zz}(\omega)$ and $-$Im$\sigma_{zz}(\omega)$ have maxima at $\omega\sim
1/\tau_z$.  Thus, although bulk disorder destroys localization at $\omega =0$, the
resulting state still has properties interpolating between those of a metal and
an Anderson insulator. This prediction is amenable to a direct experimental
verification.

Finally,
we notice that the predictions of our model are equally well applicable to a
two-dimensional (2D) case, e.g, for line barriers crossing the plane. Such a
system can be realized in a 2D electron gas with an array of randomly spaced stripe-like gates.

In conclusion, we have shown that a system with two types of disorder--randomly spaced planar barriers and bulk impurities--exhibits quite unusual transport properties. In the absence of bulk disorder, it behaves as a 1D insulator
in the out-of-plane direction and as an ideal metal in the in-plane direction. Bulk disorder renders both conductivities finite; however, $\sigma_{zz}$ \emph{increases} with bulk disorder until two disorders become comparable. For weak bulk disorder, the ratio of the conductivities may exceed the ratio of the effective masses by orders of magnitude. The \emph{ac} out-of-plane conductivity has a manifestly non-Drude frequency dependence
with a maximum at intermediate frequencies. 

We thank S. Blundell, H. Bouchiat, S. Brazovskii, K. Efetov,  A. Hebard, S. Gueron, D. Gutman, N.
Kirova, I. Lerner, G. Montambaux,  H. Pal, P. Hirschfeld, {\'E}. Rashba, A. Schofield, S. Tongay, and I. Yurkevich for
stimulating discussions. D.L.M. acknowledges the financial support from RTRA Triangle de
la Physique and hospitality of the Laboratoire de Physique des Solides,
Universit{\'e} Paris-Sud,
where a part of this work was done.
D.L.M. and V.I.Y. acknowledge hospitality of ICTP (Trieste). 
V.I.Y. acknowledges RFBR grant 09-02-01235. A.M.S and M.O. acknowledge financial
support from the Spanish DGI, project FIS2006-11126, and Fundacion Seneca, project
08832/PI/08.


\begin{thebibliography}{99}
\bibitem{comment} A further increase in bulk disorder will lead eventually
to a 3D Anderson localization, but we are not studying this regime here.
\end{thebibliography}
\end{document}